\documentclass{article}

\usepackage{spconf,amsmath,epsfig}
\usepackage{amssymb}
\usepackage{setspace}
\usepackage[svgnames]{xcolor}
\usepackage{graphics,graphicx}
\graphicspath{{./figures/}}
\usepackage{physics}
\usepackage[hidelinks]{hyperref}
\usepackage{wasysym}
\renewcommand{\i}{\textrm{i}}
\usepackage{booktabs}
\usepackage{caption}
\usepackage{enumitem}

\usepackage{siunitx}
\title{\uppercase{Improved Observation of Transient Phenomena with Doppler Radars: a Common Framework for Oceanic and Atmospheric Sensing}}

\twoauthors
  {Baptiste Domps, Julien Marmain}
	{Degreane Horizon\\
	Radar Department\\ 
	Cuers, France\\
	baptiste.domps@degreane-horizon.fr}
  {Charles-Antoine Gu\'erin}
	{Universit\'e de Toulon, Aix-Marseille Univ.,\\CNRS, IRD, MIO\\
	Toulon, France\\
	guerin@univ-tln.fr}

\begin{document}

\maketitle

\begin{abstract}
	Doppler radars are routinely used for the remote sensing of oceanic surface currents and atmospheric wind profiles. Even though they operate at different frequencies and address different media, they follow very similar processing for the extraction of measured velocities. In particular they both face the challenging issue of capturing geophysical phenomena which vary rapidly with respect to the typical integration time. Recently, the authors applied a non-spectral formalism based on autoregressive processes to model the backscattered time series obtained from High-Frequency oceanic radars. They showed that it allows to calculate Doppler spectra for very short integration times without losing in frequency resolution nor signal-to-noise ratio. We apply this technique to synthetic and experimental data within a common framework and show for the first time the strong potential of the method for the study of transient atmospheric phenomena.
\end{abstract}


\section{Introduction}

Doppler radars have been customarily used for decades for measuring wind profiles in the air column as well as oceanic currents at the sea surface (e.g \cite{liu2019,roarty2019}). Even though the physical mechanisms driving the backscattering from atmospheric and oceanic media are very different, there are many formal analogies in the description of the received time signal and its conversion to geophysical variables. In both cases the derivation of a radial velocity, which can be further interpreted in terms of wind speed or surface current, relies on measuring a Doppler shift with respect to some reference frequency, namely the zero Doppler in the atmospheric case and the Bragg frequency in the oceanic case. In either situation the accuracy of the measurement is limited by the coherent observation time which is necessary to produce a Doppler spectrum. As it is well known, the choice of the observation time results from a trade-off between the required duration for sufficient Doppler frequency resolution and Signal-to-Noise Ratio (SNR) and the maximum time scale over which the geophysical observables can be assumed stationary. The typically employed observation times are of the order of a few tens of seconds for VHF/UHF radar Wind Profilers (WP) and a few tens of minutes for oceanographic High-Frequency Radars (HFR).
%\begin{table}[h]
%	\centering
%	\caption{Radar parameters used in this study.}
%	\small
%	\begin{tabular}{lcc}
%		\toprule
%		\null & PCL-1300 WP & WERA HFR \\ \midrule
%		Modulation & Pulsed & FMCW \\
%		Central Frequency & \SI{1.274}{\giga\hertz} & \SI{13.5}{\mega\hertz} \\
%		%Resonant Length $\Lambda$ & $\approx\SI{11.5}{\cm}$ & $\approx\SI{10}{\m}$ \\
%		Pulse/Chirp Length $\Delta t$ & \SI{2.5}{\micro\s} & \SI{0.26}{\s} \\
%		Interpulse Period & \SI{70}{\micro\s} & N/A \\
%		%Range Resolution & & \\ \midrule
%		Typical Int. Time & \SI{10}{\s} & \SI{20}{min} \\
%		TVAR-MEM Int. Time & \SI{0.15}{\s} & \SI{30}{\s} \\
%		\bottomrule
%	\end{tabular}
%	\label{tab:radar}
%\end{table}
This is satisfactory for the vast majority of situations where the main atmospheric and oceanic features are only slowly varying with respect to the temporal scale of observation. However, there are some specific instances which do not comply to this observation scheme. This is the case whenever transient phenomena or rapidly evolving fields of velocities are involved, such as e.g. 1)~wind gusts, storm surges or tsunamis in the oceanic context; 2)~landing planes, bird swarms and wake turbulence echoes in the atmospheric context. This calls for specific processing of the time echo to overcome the classical time-frequency dilemma.

The authors recently applied a non-spectral, parametric approach, referred to as the Time-Varying Autoregressive Maximum Entropy Method (TVAR-MEM) to process rapidly changing oceanic data \cite{joe:domps2021,grsl:domps2021}. It is based on an Auto-Regressive (AR) representation of the received time series that allows maintaining high Doppler resolution and elevated SNR even with short samples. Due to the similarity of the scattering formalism for oceanic and atmospheric sensing (Section \ref{sec:backscattering}), the method can be also employed for atmospheric sensing and we present here its first utilization in this context. We illustrate the performances of this analysis with synthetic (Section \ref{sec:synthetic}) as well as original experimental data (Section \ref{sec:real}). We provide high-resolution Time-Frequency imaging of the radar time series that can capture some hitherto hidden signatures of birds and planes echoes.

\section{Theoretical background}\label{sec:backscattering}

As it is well known, the backscattered time series $s(t)$ from an atmospheric turbulent layer and from the sea surface share the same remarkable property, once resolved in direction: within a single-scattering approximation they are proportional to the spatial Fourier Transform of the perturbating quantity $X(\vb r,t)$ in the medium:
\begin{equation}
	\label{eq:born}
	s(t) \sim \int_\textrm{medium} X(\vb{r},t) e^{-2\i \vb{K_0}\cdot\vb{r}} \, d\vb{r}
\end{equation}
In the former case, this is obtained with the Born approximation for weak permittivity contrast (e.g. \cite{tatarskii1961}), $X$ is a contrast induced by the atmospheric particles and $\vb K_0$ is the (three-dimensional) incident EM wave vector; in the latter case, this results from the perturbation theory for shallow rough surfaces (e.g. \cite{elfouhaily2004}), $X$ is the contrast of elevation induced by waves at the sea surface and $\vb K_0$ is the (two-dimensional) horizontal projection of the incident EM wave vector. In both cases, the backscattering echo is mainly caused by resonant structures having a typical length comparable to half the radar wavelength, a result known as ``Bragg law''. For clear-air scattering, such structures are ``blobs'' of turbulent air moving with the wind. They are seen in the Doppler spectrum as a single broad peak around the central Doppler shift $f_c$ induced by the radial wind speed $U_r=-\lambda f_c/2$. For a clean sea surface observed with an coastal oceanographic radar, the resonant features are the so-called Bragg waves \cite{barrick1972} which are the gravity waves at half the radar wavelength. As they can be possibly propagating in two opposite directions, the resulting Doppler spectrum generally exhibits 2 Doppler peaks $f_c^\pm=\pm f_B$ at the so-called Bragg frequency $f_B=\sqrt{g/(\pi\lambda)}$ and its opposite. Any additional surface current $U_r$ translates the 2 Bragg peaks by the same shift $-2U_r/\lambda$, so that the latter can be inverted from the residual Doppler shift.%, $U_r=-\lambda (f_c^\pm\pm f_B)/2$.

Digital computation of the Doppler spectrum is routinely achieved from the range-resolved complex voltage time series $s(t)$ using a Fast Fourier Transform (FFT) algorithm. Best frequency resolution and SNR are thus obtained for ``long'' integration times. Inversely, short integration times, such those needed to observe transient phenomena, strongly deteriorates the quality of the spectrum and eventually the Doppler estimate. Here, we use the TVAR-MEM approach \cite{grsl:domps2021} to model the backscattered Doppler spectrum at high temporal and frequency resolution. The full time series are splitted in sequences of $N$ samples, overlapping by half of their length. Each sequence is then modeled as an autoregressive (AR) process of order $p$ \cite{stoica2005}:
\begin{equation}
	\label{eq:ar}
	s(n\Delta t) = -\sum_{k=1}^pa_ks\big((n-k)\Delta t\big)+\varepsilon_n
\end{equation}
where $a_k$ are the modeling AR coefficients and $\varepsilon_n$ is a white noise. In the context of oceanographic measurements, the authors experimentally demonstrated that the best choice for the AR order $p$ is $N/2$ \cite{joe:domps2021}, a criteria we extend to the context of atmospheric analysis. The AR coefficients are here evaluated using the Maximum Entropy Method (MEM) or ``Burg method'' \cite{burg1975}, which was found efficient for short integration times. The Power Spectral Density (PSD) is finally computed from the AR coefficients:
\begin{equation}
	\label{eq:ar_psd}
	P_{AR}(\omega) = P_\varepsilon\left|1+\sum_{k=1}^pa_ke^{-ik\omega\Delta t}\right|^{-2}
\end{equation}
The fast updating of the AR coefficients makes them ``time-varying'' (TV) and we will refer to this method as TVAR-MEM. The temporal fluctuations of the backscattered Doppler spectrum, evaluated at rapid scale with the TVAR-MEM, can visually be assessed by representing the PSD in the Time-Doppler plane. We will further refer to this representation as the ``Time-Doppler spectrogram''. Despite being commonly used in the radar community, this representation has found little to none applications to HFR nor WP until now, because of the ``long'' integration times usually required. The TVAR-MEM approach alleviates this issue.

\section{Assessment with Synthetic Data}\label{sec:synthetic}

\begin{figure}[h]
    \vspace{-2mm} % PAS BIEN
	\centering
	\includegraphics[width=\linewidth]{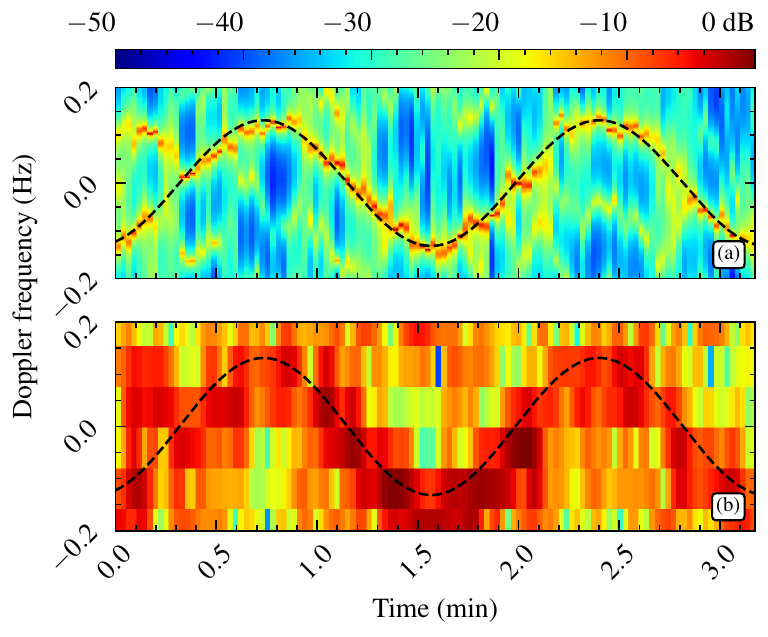}
	\vspace{-7mm}
	\caption{Simulation of the normalized PSD (colorscale; \si{\decibel}) that would obtained with a HFR. The representation is in the Time-Doppler plane, where the vertical axis is the residual Doppler frequency. The instantaneous PSD is computed from overlapping synthetic time series of $N=$\,\SI{128}{points} (\SI{33}{\s}) in presence of a rapidly-varying radial surface current $U_r(t)$ and limited to the positive Bragg line: (a)~TVAR-MEM; (b)~FFT. Simulated Doppler shift $f_D$ is superimposed as dashed line.}
	\label{fig:td_simu}
\end{figure}

We will first assess the performances of the TVAR approach in a common formalism including both atmospheric and oceanic remote sensing. For this, we simulate radar time series following the approach proposed by \cite{zrnic1975}. A typical backscattered Doppler spectrum $P$ can be written as $P(\omega) = -\big(S(\omega)+N\big)X(\omega)$
%\begin{equation}
%	\label{eq:dopplerSpectrum}
%	P(\omega) = -\big(S(\omega)+N\big)X(\omega)
%\end{equation}
where $S$ is the signal PSD, $N$ is the uniform PSD of white noise and $X$ is an exponentially distributed random variable. The complex voltage time series can then be obtained (up to a scaling factor) with a Discrete Fourier Transform of the complex spectral components:
\begin{equation}
	\label{eq:simu}
	s(t) \sim \sum_{j} \sqrt{P(\omega_j)}e^{\i\big(\omega_j t+\varphi_j\big)}e^{\mathrm{i}\Phi_D(t)}
\end{equation}
where $\varphi_j$ are uniform independent random phases. By construction, the amplitudes $\sqrt{P(\omega_j)}$ are Rayleigh distributed and the individual frequency components are complex Gaussian variables. The deterministic varying phase $\Phi_D(t)$ represents the phase shift induced by the velocity of perturbations, $\Phi_D(t) = \int_0^tU_r(\tau)\,d\tau$. In the oceanic context, $U_r(\tau)$ is the instantaneous radial surface current and the integral $\Phi_D(t)$ is referred to as the ``Memory Term'', see e.g. \cite{guerin2018}; in the atmospheric context, $U_r(\tau)$ is the radial wind speed. The memory term accounts for the possible fluctuations of the velocity $U_r$ during the integration time and reduces to the classical Doppler shift, $\Phi_D(t)=4\pi/\lambda U_r t=\omega_D t$, whenever the velocity can be assumed constant over the integration time. The resonant frequency peaks in the signal PSD $S$ are modeled with a pair of Gaussian functions centered at plus or minus the Bragg frequency (HFR) or a single Gaussian shape centered at the null Doppler frequency (WP). As a generic example we have generated a backscattered time series corresponding to \SI{20}{\mega\hertz} radar carrier frequency at a sampling rate $\Delta t=$\,\SI{100}{\milli\s}. A single positive Bragg line of width $\sigma=$\,\SI{3e-3}{\m\per\s} has been assumed with rapidly varying velocity $U_r(t) = U_0\cos\left(\omega_0 t\right)$ where $U_0=$\,\SI{0.1}{\m\per\s} and $\omega_0=$\,\SI{6e-2}{\radian\per\s}. The instantaneous PSD has been recalculated from the time series using either the TVAR-MEM or the classical FFT approach by processing half overlapping series of $N=$\,\SI{128}{samples} (i.e. \SI{33}{\s}). The chosen values correspond to the typical case of a HFR observing surface currents but could be simply rescaled to be consistent with the case of a WP sensing wind velocity. Figure \ref{fig:td_simu} shows the Time-Doppler spectrograms obtained with the two methods. The temporal variations of the Bragg line are accurately rendered with the TVAR-MEM, while barely visible using FFT.

\vspace{-.3cm} % PAS BIEN

\section{Application to Experimental Data}\label{sec:real}

Next we present an application of the TVAR-MEM approach to two experimental data sets. The first has been routinely acquired by the WERA HFR (Helzel GmbH) of Tofino, on the Pacific Coast of Vancouver Island, British Columbia; the selected time series has been recorded during the passage of a an abnormal transitory oceanic and atmospheric event. The second has been acquired with the Degreane Horizon PCL-1300 WP during the SESAR experiment that took place near the landing runways of Paris Charles de Gaulle Airport.

\subsection{The October 2016 ``Meteotsunami'' in Tofino}

\begin{figure}[h]
	\centering
	\includegraphics[width=\linewidth]{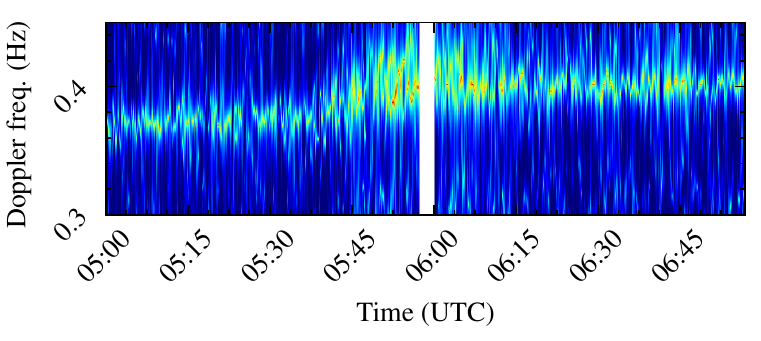}
	\vspace{-7mm}
	\caption{Normalized HFR PSD (same colorscale as Figure~\ref{fig:td_simu}; \si{\decibel}) in the Time-Doppler plane, computed from overlapping synthetic time series of $N=$\,\SI{128}{points} (\SI{33}{\s}) every $\tau=$\,\SI{16.5}{\s}, from data acquired with the HFR of Tofino on October 14, 2016 and limited to the positive Bragg line within a $f_B\,\pm\,$\SI{7.5e-3}{\hertz} window (i.e. $\pm\,$\SI{75}{\cm\per\s}). Vertical bar is an interruption in acquisition for quality control.}
	\label{fig:td_meteotsunami}
\end{figure}

On October 14, 2016, the HFR of Tofino raised a tsunami alert based on the measurements of strong abnormal surface currents. Due to the absence of any recorded seismic activity, this event was related to the family of atmospheric-induced tsunamis \cite{guerin2018,grsl:domps2021} and can be used as benchmark for tsunami detection algorithms. Here, we apply the TVAR-MEM to model the backscattered Doppler spectrum at high temporal scale. The time series $s(t)$ were processed by half overlapping blocks of $N=$\,\SI{128}{samples} (i.e. \SI{33}{\s}). Figure \ref{fig:td_meteotsunami} is the resulting Time-Doppler spectrogram centered on the positive Bragg line. The fine frequency resolution reveals a micro-Doppler jump of \SI{2.5e-2}{\hertz} (i.e. \SI{25}{\cm\per\s}) at 05:40~UTC, corresponding to a sudden surge of surface current. Furthermore, the increase of the positive Bragg line amplitude corresponds to a strengthening of the Bragg waves advancing towards the radar, confirmed by a sudden \SI{20}{\cm} rise in sea level measured by coastal tide gauges. 

\subsection{Various Transient Atmospheric Events}\label{sec:sesar}

\begin{figure*}
	\includegraphics[width=\linewidth]{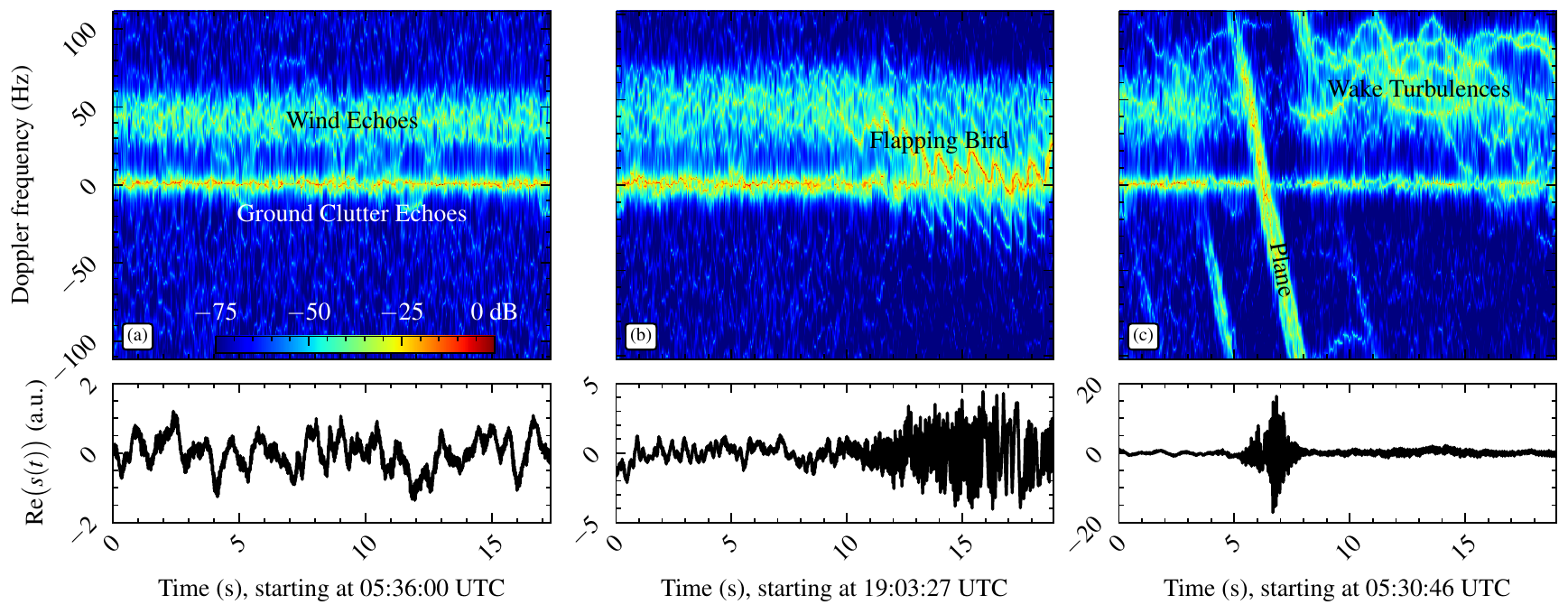}
	\vspace{-7mm}
	\caption{Data acquired with the PCL-1300 WP during the SESAR experiment. \textbf{Top:} Normalized PSD (colorscale; \si{\decibel}) in the Time-Doppler plane, computed with TVAR-MEM from overlapping samples of $N=$\,\SI{32}{points} (\SI{0.15}{\s}) every $\tau=$\,\SI{75}{\milli\s}. \textbf{Bottom:} Real part of the radar time series $s(t)$ showing the ``contamination'' by transient phenomena.}
	\label{fig:td_sesar}
\end{figure*}

The SESAR P12.2.2 XP1 campaign was conducted in autumn 2012 on the Paris CDG airport. The Degreane Horizon PCL-1300 WP was installed vertically below the ``Outer Marker'' of the landing runway 26L, which is located about \SI{10}{\km} East of the runway and marks the begin of the final approach segment. We have analyzed the complex voltage time series received on an antenna pointed towards the landing axis at a \SI{73}{\degree} site angle. The time series were processed using TVAR-MEM from overlapping blocks of $N=$\,\SI{32}{points} (\SI{0.15}{\s}) and updated every $\tau=$\,\SI{75}{\milli\s}. We selected for the illustration 3 specific events, acquired on September 24, 2012 at range gate 3 (altitudes 720 to \SI{1075}{\m}). Figure \ref{fig:td_sesar} shows the TVAR-MEM corresponding Time-Doppler spectrograms and time series.

\begin{description}[style = unboxed, leftmargin = 0cm]

	\item[(a) Wind Echoes:] Typical steady wind echoes are seen in Figure \ref{fig:td_sesar}a as the horizontal strip around the frequency $f_D=$\,\SI{45}{\hertz} (i.e. $U_r=$\,\SI{5.29}{\m\per\s}). The instantaneous PSD (vertical slices in the Time-Doppler representation) have a Gaussian shape around this central frequency. The marked horizontal line around the zero Doppler frequency corresponds to the dominant echo of fixed target.
	
	\item[(b) Flapping Bird:] Clear-air echoes are contaminated by avian echoes starting at 19:03:37~UTC (Figure \ref{fig:td_sesar}b). A bird is flying towards the radar at a radial speed varying from 4.7 to \SI{0}{\m\per\s}. The wingbeat frequency can be extracted from the micro-Doppler oscillations and is here close to \SI{3}{\hertz}.
	
	\item[(c) Plane and Wake Turbulences:] Strong echo of an airplane is located in the the first \SI{10}{\s}. Note that the airplane speed exceeds the Nyquist frequency leading to aliased echo. Assuming constant speed and altitude during the record, one can infer a Doppler rate of change of \SI{66}{\hertz\per\s}, corresponding to an average plane radial speed of \SI{75}{\m\per\s} which is consistent with the typical landing speed of commercial aircrafts. The echo is followed by multiple oscillating echoes which we attribute to wake vortex turbulence.
	
\end{description}

\section{Conclusion}

We have presented the first application of the TV-AR-MEM in the double context of ocean and atmospheric sensing. It has been applied for the first time to an experimental WP dataset. The resulting Time-Doppler maps unveil details of rapid atmospheric variations at the scale of one second, such as bird flapping or turbulence in the wake of a plane. Further work is in progress to confirm the strong potential of this technique.

\vfill{\small
%\begin{center}
%	{\bf ACKNOWLEDGMENTS\vspace{-.5em}\vspace{0pt}}
%\end{center}

\noindent \textbf{Acknowledgments:} First author was supported by the Direction G\'en\'erale de l'Arme\-ment (DGA). We are grateful to Ocean Networks Canada for providing HFR data and to Dr~Philipp Currier for countless discussions on WP.

\bibliographystyle{IEEEtran}
\bibliography{bibigarss}}

\end{document}